\documentclass[sn-mathphys]{sn-jnl}

\jyear{2021}%

\newtheorem{theorem}{Theorem}
\newtheorem{coro}{Corollary}%

\theoremstyle{thmstyletwo}%

\theoremstyle{thmstylethree}%

\usepackage{hyperref}
\usepackage{bm}
\usepackage{threeparttable}
\usepackage{multirow}
\newcommand{\bas}{\begin{eqnarray*}}
\newcommand{\eas}{\end{eqnarray*}}
\newcommand{\ba}{\begin{eqnarray}}
\newcommand{\ea}{\end{eqnarray}}
\newcommand{\pr}{{\rm pr}}

\newcommand{\bbeta}{{ \bm \beta}}

\newcommand{\etab}{{\bm \eta}}
\newcommand{\rV}{ {\rm V}}

\newcommand{\bX}{{\bf X}}
\newcommand{\bY}{{\bf  Y}}
\newcommand{\bZ}{{\bf  Z}}
\newcommand{\bA}{{\bf  A}}
\def\bx{{\bf x}}
\def\by{{\bf y}}
\newcommand{\bz}{{\bf  z}}
\newcommand{\bzero}{{\bf 0}}

\newcommand{\bV}{{\bf V}}
\newcommand{\bS}{{\bf S}}
\newcommand{\bSigma}{{\bf \Sigma}}

\newcommand{\bU}{{\bf U}}
\def\bu{{\bf u}}
\def\rv{{\rm v}}
\newcommand{\bO}{ { O}}
\usepackage{pmat}
\usepackage{arydshln}
\def\T{{ \mathrm{\scriptscriptstyle T} }}
\usepackage[none]{hyphenat}
\usepackage{natbib}

\begin{document}
\sloppy
\title[Two-step empirical likelihood inference from capture--recapture data]
{Two-step semiparametric empirical likelihood inference from capture--recapture data with missing covariates}


\author[1]{\fnm{Yang} \sur{Liu}}\email{liuyangecnu@163.com}

\author*[1]{\fnm{Yukun} \sur{Liu}}\email{ykliu@sfs.ecnu.edu.cn}

\author[2]{\fnm{Pengfei} \sur{Li}}\email{pengfei.li@uwaterloo.ca}

\author[3]{\fnm{Riquan} \sur{Zhang}}\email{zhangriquan@163.com}

\affil[1]{\orgdiv{KLATASDS -- MOE, School of Statistics}, \orgname{East China Normal University}, \orgaddress{
            \city{Shanghai}, \postcode{200062}, \state{Shanghai},
            \country{China}}}

\affil[2]{\orgdiv{Department of Statistics and Actuarial Science}, \orgname{University of Waterloo}, \orgaddress{
\city{Waterloo}, \postcode{N2L 3G1}, \state{Ontario}, \country{Canada}}}

\affil[3]{\orgdiv{School of Statistics and Information},
\orgname{Shanghai University of International Business and Economics},
\orgaddress{
\city{Shanghai}, \postcode{201620}, \state{Shanghai}, \country{China}}}


\abstract{Missing covariates are not uncommon in capture--recapture studies.
When covariate information is missing at random in capture--recapture data,
an empirical full likelihood method has been demonstrated to outperform conditional-likelihood-based methods in abundance estimation.
However, the fully observed covariates must be discrete, and the method is not directly applicable to continuous-time capture--recapture data.
Based on the  Binomial  and Poisson regression models, we propose a two-step
semiparametric empirical likelihood approach
for abundance estimation in the presence of missing covariates,
regardless of whether the fully observed covariates are discrete or continuous.
We show that the maximum semiparametric empirical likelihood estimators
for the underlying parameters and the abundance are
 asymptotically normal, and more efficient
than the counterpart for a completely known non-missingness probability.
After scaling, the empirical likelihood ratio test statistic for
 abundance follows a limiting chi-square distribution with one degree of freedom.
The proposed approach is further extended to one-inflated count regression models,
and a score-like test is constructed to assess whether one-inflation exists among the number of captures.
Our simulation shows that, compared with the previous method, the proposed method not only performs better in correcting bias, but also has a more accurate coverage in the presence
of fully observed continuous covariates, although there may be a slight efficiency loss when the
fully observed covariates are only discrete.
The performance of the new method is illustrated by an analysis of the Hong Kong prinia data.}




\maketitle

\section{Introduction}
\label{s:intro}

As cost-effective sampling techniques, capture--recapture experiments
have been widely used to infer the size or abundance of a
finite population, such as a species,
or a hidden population \cite{borchers2002estimating, mccrea2014analysis,bohning2017capture}.
We assume that the population is closed,
namely there are no births, deaths, or migration,
so that the abundance remains unchanged.
During an experiment, individual covariates are frequently recorded
together with capture histories,
and they are used to model the probability of an individual being captured.
However, some individual covariates may be subject to missingness
due to improper measurement or other reasons.
According to \cite{little2014statistical} and \cite{rubin1976inference},
such covariates are (1) missing completely at random,
if the missingness probability
does not depend on any data,
(2) missing at random, if the missingness probability
does not depend on the unobserved data
conditional on the observed data,
or (3) missing not at random, if the missingness probability
still depends on the unobserved data themselves
even conditional on the observed data.

A large proportion of data that are missing completely at random
can be attributed to careless operation or ignorance
about the process of data collection.
In this situation,
the abundance can be inferred directly from the complete cases.
To address the missing data and measurement errors simultaneously,
\cite{xi2009estimation} constructed
a set of estimating equations and
proposed an expectation--maximization-type algorithm for abundance estimation.
The missing-not-at-random mechanism means that
the missingness probability of a covariate depends on itself.
For example, female animals may be more likely to be missing
than male animals, if, in nature, the males are easier to identify.
For such data, \cite{liu2016population} and
\cite{liu2020abundance}, respectively, developed
conditional and full likelihood abundance estimation methods by
building a sub-model for the non-missingness probability.

Compared with the missing-completely-at-random and
missing-not-at-random mechanisms,
the missing-at-random mechanism is more frequently encountered in
capture--recapture experiments.
As measurements are usually taken each time an individual is captured,
it is generally true that the fewer the number of captures,
the higher the probability of covariates being missing.
Quite a few estimation methods have been developed
to handle capture--recapture data with missing-at-random covariates.
\cite{lee2016estimation} proposed that first, the regression parameters should be estimated
by the inverse probability weighting and multiple imputation methods,
 and then several Horvitz--Thompson-type estimators should be constructed for the abundance.
These methods were extended by \cite{stoklosa2019closed} and
\cite{noghrehchi2020multiple}
to more general cases with both missing data and measurement errors.
Based on conditional likelihood,
\cite{liu2019abundance} constructed an
optimal estimating function to estimate
the regression parameters.
By combining conditional likelihood and empirical likelihood,
\cite{liu2020maximum} proposed a
maximum empirical likelihood abundance estimation approach,
which seems more attractive than conditional-likelihood-based methods.
The maximum empirical likelihood estimator of the abundance
has a smaller mean square error than
the Horvitz--Thompson-type estimators, and
the empirical likelihood ratio confidence interval has a
more accurate coverage probability than
conditional-likelihood-based Wald-type confidence intervals.

In this article, we consider only the case where
the covariates are missing at random.
\cite{liu2020maximum}'s maximum likelihood estimation method requires
the fully observed covariates to be discrete and is, therefore, not directly applicable if some fully observed covariates have continuous values.
An example is wing length in the Hong Kong prinia data,
which is analysed in Section~\ref{s:data}.
We could transform the fully observed continuous covariates
into discrete variables and then apply \cite{liu2020maximum}'s method.
However, this transformation may lead to information loss, and
the new variables may distort the relation between the covariates
and the capture probability.
The resulting maximum empirical likelihood estimator may
have a large bias and the resulting
empirical likelihood ratio confidence interval may have severe undercoverage.
In addition, \cite{liu2020maximum}'s method
is not directly applicable to continuous-time
capture--recapture data as it is designed for
discrete-time capture--recapture data.

Under the  Binomial  and Poisson regression models
for discrete- and continuous-time capture--recapture data, respectively,
we propose a two-step semiparametric empirical likelihood
approach to abundance estimation.
In the first step, we fit the fully observed data by
a parametric non-missingness probability model.
In the second step, we construct a semiparametric empirical likelihood
 by combining an empirical likelihood and a partial likelihood,
with the non-missingness probability model replaced by its
estimator obtained in the first step.
The abundance and the other unknown parameters are estimated by
maximizing the resulting two-step maximum empirical likelihood.
Theoretically, we prove that the estimator is asymptotically normal
and more efficient than the counterpart if there is a known non-missingness probability.
We also show that, after appropriate scaling,
the empirical likelihood ratio test statistic for abundance
has a limiting chi-square distribution with one degree of freedom.
Compared with \cite{liu2020maximum}'s method, our simulation shows that the proposed two-step empirical likelihood method is not only better at correcting bias,
but also produces more accurate confidence intervals
in the presence of fully observed continuous covariates,
although there may be a slight loss of efficiency if all the fully observed covariates are discrete.

One-inflation problems have recently received much attention
in the context of capture--recapture models.
\cite{godwin2017estimation} first noticed
that the number of captures usually exhibits a preponderance of 1-counts
and that ignoring the excess 1's may result in upwards biased estimates of
population sizes.
In the absence of missing data,
many one-inflated count regression models have been proposed that account for the one-inflation problem. See
\cite{bohning2021population, bohning2020general, bohning2019identity, godwin2017one, godwin2019one, godwin2017estimation}
and references therein.
To the best of our knowledge, few methods in the literature can cope with the one-inflation problem in the presence of missing data.
As a complement, we extend the two-step empirical likelihood estimation approach
to deal with one-inflated capture--recapture data with missing covariates.
To test whether there is one-inflation among the number of captures,
a score-like test is proposed, and its limiting distribution is also derived.

The remainder of this article is organized as follows.
In Section \ref{s:like},
we introduce the  capture-recapture data and models,
and develop the two-step semiparametric empirical likelihood method for abundance estimation.
The asymptotic distributions of the
maximum empirical likelihood estimator and the empirical likelihood ratio statistic are established,
and further extended to the one-inflated capture-recapture models.
 Section \ref{s:sim}  demonstrates the finite-sample performance of
the proposed methods through several simulation studies.
A yellow bellied prinia data from Hong Kong is analyzed
in Section \ref{s:data} to illustrate
the proposed two-step empirical likelihood based methods.
 Section \ref{s:con} concludes with a discussion.
For the convenience of presentation,
all proofs are given in the online supplementary information.

\section{Semiparametric empirical likelihood inference}
\label{s:like}

\subsection{Model and data}
\label{sec:mod-lik}
Let $N $ be the abundance or size of the finite population of interest.
Suppose that a capture--recapture experiment
is conducted over $K$ capture occasions
or during a period of time.
Let $D$ be the number of times that a generic individual in the population
has been captured during the experiment.
We use $\bZ = (1, \bX^\T, \bY^\T)^\T$ to denote the covariate of an individual,
where $\bX$ is fully observed and $\bY$ is subject to missingness.
Individuals usually behave independently across occasions in capture--recapture experiments.
We assume that $D$ given $\bZ = \bz$ follows a  Binomial  distribution
in discrete-time experiments and a Poisson distribution
in continuous-time experiments.
Specifically, the probability mass function $P(D=k\mid \bZ=\bz)$
is modelled by
\ba\label{eq:mod-bi}
f(k, \bz;\bbeta) \!=\! {K \choose k} \{g(\bz; \bbeta)\}^{k} \{1 - g(\bz; \bbeta)\}^{K- k},\;
g(\bz; \bbeta) \!=\! \frac{\exp(\bbeta^\T \bz)}{1 + \exp(\bbeta^\T \bz)},
\ea
for $k= 0, \ldots,K$ and $f(k, \bz;\bbeta) = 0$ for $k=K+1,\dots$,
or
\ba\label{eq:mod-po}
f(k, \bz; \bbeta) \!=\!
\frac{\{\lambda(\bz; \bbeta)\}^k}{k!}
\exp\{ - \lambda(\bz; \bbeta)\}, \;
\lambda(\bz; \bbeta) \!=\!
\exp(\bbeta^\T \bz), \;
k=0, 1,\dots.
\ea
Model~\eqref{eq:mod-bi} is the popular Huggins--Alho model
 \cite{huggins1989statistical,alho1990logistic}
and model~\eqref{eq:mod-po} is a special case of the Andersen--Gill model
with constant intensity function \cite{andersen1982cox}.
As discussed in \cite{liu2018full},
the empirical likelihood inference methods produce the same abundance estimator
under model~\eqref{eq:mod-po} as under the Andersen--Gill model.

We denote the missingness indicator of $\bY$ by $R$.
It is 1 if $\bY$ is not missing and 0 otherwise.
Suppose that $n$ individuals have been captured at least once
in the capture--recapture experiment and
that the first $m$ observations are completely observed and
the last $n - m$ observations have no values of $\bY$.
Let $\bx_i, \by_i$, $d_i$, and $r_i$
($i=1,\dots,n$) be realizations of
$\bX$, $\bY$, $D$, and $R$ given $D>0$.
The observed data can be written as $\bO_1\cup\bO_2$, where
\bas
\bO_1 \!=\! \{(d_i, \!\bx_i, \!\by_i, \!r_i=1): \!i \!=\! 1,\dots,m\},
\quad
\bO_2 \!=\! \{(d_i,\! \bx_i,\! r_i=0):\! i \!=\! m+1,\dots,n\}.
\eas
We assume that $\bY$ is missing at random,
i.e.,
\ba\label{eq:MAR-model}
P(R = 1\mid \bX=\bx, \bY = \by, D=k ) = P(R = 1\mid \bX=\bx, D=k)
 =:\pi(\bx, k).
\ea
When the components of $\bX$ are all categorical,
\cite{liu2020maximum} proposed a
maximum likelihood approach to abundance estimation based on
discrete-time capture--recapture data.
However, this approach is not directly applicable to
the cases where some components of $\bX$ are continuous
or the capture--recapture data come from a continuous-time experiment.

\subsection{Semiparametric empirical likelihood}
\label{s:sel}

Our proposed estimation method depends mainly on
the full likelihood $L$ based on the complete-case data $\bO_1$.
We write $L = L_1\times L_2$, where $L_1$ is the probability of the event that
$m$ individuals have been captured at least once
and their covariates are completely observed.
 $L_2$ is the probability of the observations
$O_1$, conditional on the foregoing event.

Let $\alpha = P( D > 0, R=1)$.
As $m$ follows a  Binomial  distribution Bi$(N, \alpha)$, then
\ba\label{eq-prob-m}
L_1 = {N \choose m} \alpha^m (1-\alpha)^{N-m}.
\ea
Let $\bz_i = (1, \bx_i^\T, \by_i^\T)^\T$ for $i=1, \dots, m$.
By Bayes' rule, $L_2$ can be expressed as
\ba\label{eq-prob-others}\notag
L_2 & =&
\prod_{i=1}^m
P( D = d_i, \bX = \bx_i, \bY = \by_i \mid D>0, R=1) \\
 & =& \frac{1}{\alpha^m} \prod_{i=1}^m
P(R=1 \mid \bZ = \bz_i, D=d_i)\
P( D=d_i \mid \bZ = \bz_i)\
P(\bZ = \bz_i).
\ea
Combining Equations~\eqref{eq-prob-m} and~\eqref{eq-prob-others} together, we have
the log-likelihood:
\bas
\log(L) &=& \log {N \choose m} + (N - m) \log(1 - \alpha) +
\sum_{i = 1}^m \log\{f(d_i, \bz_{i};\bbeta)\}
\\
&&\qquad +
\sum_{i = 1}^m \log\{\pi(\bx_i, d_i)\} +
\sum_{i = 1}^m \log\{P(\bZ = \bz_i)\},
\eas
where we have used models~\eqref{eq:mod-bi}--\eqref{eq:MAR-model}.
Although the function $\pi(\bx, d)$ is unknown,
we proceed as if it were known.
Its estimation will be studied in Section~\ref{sec:step-one}.

By the principle of empirical likelihood
\cite{owen1988empirical,owen1990empirical},
we assume that the cumulative distribution of $\bZ$ satisfies
$
F(\bz) = \sum_{i=1}^m p_i I(\bz_i\leq \bz),
$ where
$p_i \geq 0$ for $i=1,\dots,m$ and
$\sum_{i=1}^m p_i=1$.
We define
$
\phi(\bz; \bbeta)=
\sum_{k=1}^\infty
\pi(\bx, k) f(k, \bz; \bbeta),
$
which is equal to $P(D>0, R=1 \mid \bZ = \bz)$.
Because
\bas
\alpha = P(D>0, R=1 )
= \e \{ P(D>0, R=1 \mid \bZ )\} = \e\{\phi(\bZ; \bbeta)\},
\eas
then feasible $ p_i$'s should satisfy:
\bas
\sum_{i=1}^m \left\{
\phi(\bz_i; \bbeta)
-\alpha\right\} p_i = 0.
\eas
Substituting $P(\bZ = \bz_i) = p_i$ into
$\log(L)$ and
applying the method of Lagrange multipliers, we find that
the maximum of $\log(L)$ is at
\bas
p_i = \frac{1}{m} \frac{1}{1 + \xi\{\phi(\bz_i; \bbeta) - \alpha\}},
\quad i = 1,\dots, m,
\eas
where $\xi = \xi(\bbeta, \alpha)$ satisfies
\bas
\sum_{i = 1}^m \frac{\phi(\bz_i; \bbeta) - \alpha}
{1 + \xi\{\phi(\bz_i; \bbeta) - \alpha\}} = 0.
\eas
Accordingly, we have the profile empirical log-likelihood of $(N, \bbeta, \alpha, \pi)$ up to a constant:
\ba
\label{eq:log-EL}\notag
\ell(N, \bbeta, \alpha, \pi)& = &\log \binom{N}{m} + (N - m) \log(1 - \alpha)
 - \sum_{i = 1}^m \log[1 + \xi\{\phi(\bz_i; \bbeta) - \alpha\}] \\
 &&\quad + \sum_{i = 1}^m [\log\{f(d_i, \bz_{i};\bbeta)\} + \log\{\pi(\bx_i, d_i)\} ].
\ea

The above form of the log-likelihood is not suitable for practical use
because the function $\pi(\bx, k)$ is unknown in general.
To overcome this problem, we propose a two-step estimation procedure.
In the first step,
we estimate $\pi(\bx, k)$ under a parametric model assumption (Section~\ref{sec:step-one}).
In the second step, putting the estimated $\pi(\bx, k)$ into $\ell(N, \bbeta, \alpha, \pi)$
gives a practically available semiparametric empirical likelihood function,
and we propose to estimate $N$ by maximizing the resulting
empirical likelihood function.
To be distinct from \cite{liu2020maximum}'s method,
the proposed method is termed the two-step semiparametric empirical likelihood method.

\subsection{Estimating $\pi(\bx, k)$}
\label{sec:step-one}

We assume a parametric non-missingness probability model:
\ba\label{eq:sel-prob-para}
\pi(\bx, k; \etab) = \frac{\exp\{(1, \bx^\T, k) \etab\}}
{1 + \exp\{(1, \bx^\T, k) \etab\}},\quad
k=1, 2, \dots,
\ea
where $\etab$ is an unknown parameter vector.
Because $\bX$ and $D$ are fully observed,
$\etab$ can be inferred from
the fully-observed data $\{(\bx_i, d_i, r_i): i = 1,\dots,n\}$.
Given $\{(\bx_i, d_i): i = 1,\dots,n\}$,
the conditional log-likelihood based on $\{r_i=1, i =1,\dots, m\}
\cup\{r_i = 0, i =m+1,\dots, n\}$ is
\bas
\ell_\pi(\etab) = \sum_{i=1}^m \log\{\pi(\bx_i, d_i; \etab)\} +
\sum_{i=m+1}^n \log\{1-\pi(\bx_i, d_i; \etab)\}.
\eas
We denote the maximum likelihood estimator of $\etab$
by $\widehat\etab = \arg\max\{\ell_\pi(\etab)\}$.
The following theorem establishes its asymptotic normality.

\begin{theorem}\label{prop:eta}
Let $N_0$ and $\etab_0$ be the true values of $N$ and
$\etab$, respectively. We define
\bas
\bU =
\mathbb{E}\left\{
I(D>0) \pi(\bX, D; \etab_0)
\{1-\pi(\bX, D; \etab_0)\}
\begin{pmat}[{.}]
1\cr
\bX\cr
D\cr
\end{pmat}^{\otimes2}
\right\},
\eas
where
${ \bA}^{\otimes2} = \bA \bA^\T$ for a vector $\bA$.
If the non-missingness probability model in \eqref{eq:sel-prob-para} is true and
$\bU$ is positive definite, then as $N_0\to\infty$, the distribution  of $
N_0^{1/2} (\widehat \etab - \etab_0)
$
converges to the normal distribution with mean zero and covariance matrix $\bU^{-1}$.
\end{theorem}

\subsection{Two-step semiparametric empirical likelihood}
\label{sec:two-step-EL}

With $\pi(\bx, k; \widehat\etab)$ in place of $\pi(\bx, k)$
in the profile empirical log-likelihood~\eqref{eq:log-EL},
we have a practically useful log-likelihood function:
\bas
\ell(N, \bbeta, \alpha) &=& \log \binom{N}{m} + (N - m) \log(1 - \alpha)
 + \sum_{i = 1}^m \log\{f(d_i, \bz_i; \bbeta)\}
 \\
&&\qquad - \sum_{i = 1}^m \log[1 + \xi\{\phi(\bz_i; \bbeta, \widehat\etab) - \alpha\}],
\eas
where $\xi = \xi(\bbeta, \alpha, \widehat\etab)$ satisfies
\bas
\sum_{i = 1}^m \frac{\phi(\bz_i; \bbeta, \widehat\etab) - \alpha}
{1 + \xi\{\phi(\bz_i; \bbeta, \widehat\etab) - \alpha\}} = 0,
\quad
\phi(\bz; \bbeta, \etab) =
\sum_{k=1}^\infty
\pi(\bx, k; \etab)f(k, \bz; \bbeta).
\eas

Based on $\ell(N, \bbeta, \alpha)$, we define the two-step maximum empirical likelihood estimator
as
$
(\widehat N, \widehat\bbeta, \widehat\alpha)
= \arg\max \{ \ell(N, \bbeta, \alpha) \}
$
and define the empirical likelihood ratio function of $N$ as
\bas
R(N) = 2\{\ell(\widehat N, \widehat\bbeta, \widehat\alpha) -
\ell(N, \widehat\bbeta_N, \widehat\alpha_N)\},
\eas
where $(\widehat\bbeta_N, \widehat\alpha_N) =
\arg\max_{(\bbeta, \alpha)} \{\ell(N, \bbeta, \alpha)\}$ for given $N$.
Theorem~\ref{thm:asy} establishes the limiting distributions
of the maximum empirical likelihood estimator and
the empirical likelihood ratio test statistic.
For ease of exposition, we define some notation.
Let $\bbeta_0$ and $\alpha_0$ be
the true values of $\bbeta$ and $\alpha$, respectively.
We define the partial derivatives
$\dot\phi_{\bbeta}(\bz;\bbeta, \etab) =
\partial\phi(\bz; \bbeta, \etab)/\partial\bbeta$,
$\dot\phi_{\etab}(\bz;\bbeta, \etab) =
\partial\phi(\bz; \bbeta, \etab)/\partial\etab$,
and
{\small
\bas
\ddot\phi_{\bbeta}(\bz;\bbeta, \etab)
\!=\! \frac{\partial\phi(\bz;\bbeta, \etab) }
{\partial \bbeta\partial \bbeta^\T},
\;
\ddot\phi_{\bbeta\etab}(\bz;\bbeta, \etab)
\!=\! \frac{\partial\phi(\bz;\bbeta, \etab) }
{\partial \bbeta\partial \etab^\T},
\;
\ddot\phi_{\etab} (\bz; \bbeta, \etab) \!=\!
\frac{\partial\phi(\bz;\bbeta, \etab) }
{\partial \etab\partial \etab^\T}.
\eas}
We use $v_f(\bz; \bbeta)$ to denote the conditional variance of
the number of captures $D$ given the covariate $\bZ = \bz$.
That is,
$v_f(\bz; \bbeta) = Kg(\bz; \bbeta)\{1 - g(\bz; \bbeta)\}$
under the  Binomial  regression model~\eqref{eq:mod-bi} and
$v_f(\bz; \bbeta) = \exp(\bbeta^\T\bz)$
under the Poisson regression model~\eqref{eq:mod-po}.
The following matrix plays a critical role in
describing the asymptotic properties of our method:
{\small
\[
\bS \!=\!
\begin{pmat}[{..}]
\bS_{11} & \bS_{12}\cr
\bS_{21} & \bS_{22}\cr
\end{pmat}
\!=\!
\left[\begin{array}{ccc;{2pt/2pt}c}
\rV_{11} & \bzero^\T & \rV_{13} & \bzero^\T\\
\bzero & \bV_{22}-\bV_{25}\rV_{55}^{-1}\bV_{52} &
\bV_{23} - \bV_{25}\rV_{55}^{-1}\rV_{53} &
\bV_{24} - \bV_{25}\rV_{55}^{-1}\bV_{54}\\
\rV_{31} & \bV_{32} - \rV_{35}\rV_{55}^{-1}\bV_{52} &
\rV_{33} - \rV_{35}\rV_{55}^{-1}\rV_{53} &
\bV_{34} - \rV_{35}\rV_{55}^{-1}\bV_{54}\\
\hdashline[2pt/2pt]
\bzero & \bV_{42} - \bV_{45}\rV_{55}^{-1}\bV_{52} &
\bV_{43} - \bV_{45}\rV_{55}^{-1}\rV_{53} &
\bV_{44} - \bV_{45}\rV_{55}^{-1}\bV_{54}
\end{array}\right],
\]}
where
$\rV_{11} = - \alpha_0/(1 - \alpha_0)$ and
other $\rV_{ij}$'s are given in Equation (4) in the supplementary information.
The submatrix $\bS_{11}$ is further partitioned as
\bas
\bS_{11}=
\begin{pmat}[{..}]
\rV_{11} & \bS_{1112}\cr
\bS_{1121} & \bS_{1122}\cr
\end{pmat}.
\eas

\begin{theorem}\label{thm:asy}
Suppose that
$\mathbb{E} (\| \bX \| )<\infty$, $\mathbb{E} (D)<\infty$,
and the matrix
$
\bSigma =- \bS_{11}^{-1} \bS_{12} \bU^{-1}
\bS_{21}\bS_{11}^{-1}
- \bS_{11}^{-1}
$
is positive definite.
Under the conditions of Theorem~\ref{prop:eta},
 as $N_0\to\infty$,
(a) the distribution of
$N_0^{1/2} \{\widehat N/N_0 - 1, (\widehat\bbeta-\bbeta_0)^\T, \widehat\alpha - \alpha_0\}^\T$
converges to the normal distribution
with mean zero and variance $\bSigma$
and
(b) the distribution of
$ R(N_0)/\{(s - \rV_{11}) \sigma^2\} $
converges to the chi-square distribution with one degree of freedom,
where $s = \bS_{1112}\bS_{1122}^{-1}\bS_{1121}$ and
$\sigma^2$ is the $(1, 1)$th element of $\bSigma$.
\end{theorem}

\begin{coro}
\label{col:discrete}
If $\pi(\bx,k)$ is replaced by
$\pi(\bx,k;\etab_0)$ in the proposed estimation procedure, the distribution of
$N_0^{1/2} \{\widehat N/N_0 - 1, (\widehat\bbeta-\bbeta_0)^\T, \widehat\alpha - \alpha_0\}^\T
$
converges to the normal distribution with mean zero
and variance $ -\bS_{11}^{-1}$.
This suggests that the proposed estimator
is asymptotically more efficient than the counterpart
for a known non-missingness probability.
\end{coro}

Reasonable estimates of the asymptotic variance
of the proposed estimator
$(\widehat N,\widehat \bbeta,\widehat \alpha)$
are needed to quantify its variability based on the observed data.
According to Theorem~\ref{thm:asy},
this is equivalent to estimating $\bSigma$ or the $\rV_{ij}$'s.
Since the $\rV_{ij}$'s can
be expressed as $\mathbb{E} \{J(\bZ; \bbeta_0, \etab_0, \alpha_0)\}$
for some function $J$,
we can estimate them with
\bas
\frac{1}{\widehat N} \sum_{i=1}^{m}
\frac{J(\bz_i; \widehat \bbeta, \widehat \etab, \widehat \alpha)}
{\phi(\bz_i; \widehat \bbeta, \widehat \etab)}.
\eas
Using the relation between $\bSigma$ and the $\rV_{ij}$'s,
accordingly, we can construct reasonable estimates for $\bSigma$ and
the asymptotic variances of $\widehat N$,
$\widehat \bbeta$, and $\widehat \alpha$.
Based on result (b) of Theorem~\ref{thm:asy},
a scaled empirical likelihood ratio confidence interval of $N$
can be constructed as
\bas
{\mathcal I}_{\rm SEL} = \{N: R(N)/
\{(\widehat s - \widehat \rV_{11})
\widehat\sigma^2\} \leq \chi^2_{1}(1 - a)\},
\eas
where
$\widehat\sigma^2$, $\widehat s$, and $\widehat \rV_{11}$
are the proposed estimators of $\sigma^2$, $s$, and $\rV_{11}$
respectively, and
$\chi^2_{1}(1 - a)$ is the $(1-a)$th quantile of $\chi^2_{1}$.
Theorem~\ref{thm:asy} implies that
$\mathcal I_{\rm SEL}$ has an asymptotically correct coverage probability
at the $(1-a)$ confidence level for any given $a \in (0, 1)$.

\subsection{An extension to one-inflated capture--recapture models}

The proposed two-step semiparametric empirical likelihood approach is very flexible. It allows a one-inflation parameter in the  Binomial  or Poisson regression model.
Given $\bZ = \bz$,
the conditional probability of $D=k$ has a closed form, i.e.,
{\small
\ba\label{eq:cap-prob-inflate}
h(k , \bz; \bbeta, \omega) =
\begin{cases}
f(0, \bz;\bbeta), 				 & k=0, \\
(1-\omega)\{1 - f(0, \bz;\bbeta)\} +
\omega f(1, \bz;\bbeta), 	 & k=1, \\
\omega f(k, \bz;\bbeta), 	 & k > 1,
\end{cases}
\ea}
where $\omega\in (0, 1]$ is an unknown one-inflation parameter.
Under this model, the conditional probability of $D = k$
given $\bZ=\bz$ and given that the individual has been captured at all is
\bas
\pr(D = k \mid \bZ = \bz, D>0) &=&
I(k=1)\left\{
(1-\omega)+
\frac{\omega f(1, \bz;\bbeta)}{1 - f(0, \bz;\bbeta)} \right\}
\\
&&\quad + I(k>1)
\frac{\omega f(k, \bz;\bbeta)}{1 - f(0, \bz;\bbeta)}.
\eas
As noted by \cite{godwin2017estimation},
the excess probability of a 1 is drawn from
the positive counts only, i.e., the inflation process
can alter non-zero counts only.
This data-generating process is aligned with the idea that
the inflation parameter $1 - \omega$ is the proportion of individuals
in the sample who gain information from the initial
capture that provides a desire and ability to avoid subsequent captures.

Using arguments like those for $\ell(N, \bbeta, \alpha)$ in Sections~\ref{s:sel}--\ref{sec:two-step-EL},
we obtain the profile empirical log-likelihood function
under the model~\eqref{eq:cap-prob-inflate}:
\bas
\ell_{e}(N, \bbeta, \omega, \alpha) &=& \log \binom{N}{m} + (N - m) \log(1 - \alpha)
 + \sum_{i = 1}^m \log\{h(d_i, \bz_i; \bbeta, \omega)\}
 \\
 &&\quad- \sum_{i = 1}^m \log[1 + \xi\{\phi_{e}(\bz_i; \bbeta, \omega, \widehat\etab) - \alpha\}],
\eas
where $\xi = \xi(\bbeta, \omega, \alpha, \widehat\etab)$ satisfies
\bas
\sum_{i = 1}^m \frac{\phi_{e}(\bz_i; \bbeta, \omega,\widehat\etab) - \alpha}
{1 + \xi\{\phi_{e}(\bz_i; \bbeta, \omega,\widehat\etab) - \alpha\}} \!=\! 0,
\;
\phi_{e}(\bz; \bbeta, \omega, \etab) \!=\! \sum_{k=1}^\infty \pi(\bx, k; \etab) h(k, \bz; \bbeta, \omega).
\eas

When $\omega = 1$,
$\ell_e(N, \bbeta, \omega, \alpha)$
 reduces to
$\ell(N, \bbeta, \alpha)$ and the corresponding maximum
empirical likelihood estimator is $(\widehat N, \widehat \bbeta, \widehat \alpha)$,
whose asymptotic property was investigated in Theorem~\ref{thm:asy}.
When $0<\omega <1$,
we define the two-step maximum empirical likelihood estimator as
$
(\widehat N_e, \widehat\bbeta_e, \widehat \omega_e, \widehat\alpha_e) =
{\arg\max} \{ \ell_e(N, \bbeta, \omega, \alpha) \}
$,
and define the empirical likelihood ratio function of $N$ as
$
R_e(N) =2\{ \ell_e(\widehat N_e, \widehat\bbeta_e,
\widehat \omega_e, \widehat\alpha_e) -
 \max_{(\bbeta, \omega, \alpha)} \ell(N, \bbeta, \omega, \alpha)\}
$.
By similar arguments as those in the proof of
Theorem~\ref{thm:asy}, we have the following theorem.

\begin{theorem}\label{the:asy-inflated}
Let $\omega_0\in(0,1)$ be the true value of $\omega$. Let
$\bSigma_e = - \bS_{11e}^{-1} - \bS_{11e}^{-1}
\bS_{12e} \bU^{-1} \bS_{21e}\bS_{11e}^{-1}$ and
$s_e = \bS_{1112e}\bS_{1122e}^{-1}\bS_{1121e}$,
where $\bS_{ije}$ and $\bS_{11ije}$ are defined in (7) of the supplementary information.
Suppose that
$\mathbb{E}(\bX)<\infty$, $\mathbb{E}(D)<\infty$,
and the matrix $\bSigma_e$
is positive definite.
As $N_0\to \infty$,
(a) the distribution of
$N_0^{1/2} \{\widehat N _e/N_0 - 1, (\widehat\bbeta _e-\bbeta_0)^\T,
\widehat \omega _e - \omega_0, \widehat\alpha _e - \alpha_0\}^\T$
converges to the normal distribution with mean zero and
variance $ \bSigma_e$ and (b) the distribution of
$\{(s_e - \rV_{11}) \sigma^2_e\}^{-1}R_e(N_0)$
converges to the chi-square distribution with one degree of freedom,
where $\sigma^2_e$ is the $(1, 1)$th element of $\bSigma_e$.
\end{theorem}

Consistent estimates of $s_e$ and $\sigma_e^2$ are needed
if we construct likelihood-ratio-based confidence intervals
following Theorem~\ref{the:asy-inflated}.
Such estimates can be constructed
by the technique of
constructing $\widehat s$ and $\widehat \sigma^2$
at the end of Section~\ref{sec:two-step-EL}.
We denote the resulting estimates by $\widehat s_e$ and $\widehat\sigma_e^2$.
Result (b) of Theorem~\ref{the:asy-inflated} implies
that a reasonable scaled empirical likelihood ratio confidence interval of $N$ at
the $(1-a)$ level is
$
\mathcal{I}_{{\rm SEL}e}
= \{N: \{(\widehat s_e - \widehat\rV_{11})
 \widehat\sigma^2_e\}^{-1} R _e(N) \leq \chi^2_1(1-a)\}.$

An interesting problem with the one-inflated capture--recapture model
is whether there is one-inflation among the number of captures,
which is equivalent to testing $H_0: \omega = 1$ under model~\eqref{eq:cap-prob-inflate}.
\cite{godwin2017estimation} investigated this problem
in the absence of missing covariates, and they recommended the use of score tests.
Taking the partial derivative of $\ell_e(N, \bbeta, \omega, \alpha)$
with respect to $\omega$ at $\omega=1$ gives the score function:
\bas
\frac{\partial \ell_e( N, \bbeta, 1, \alpha)}{\partial w}
=
\sum_{i=1}^m \left\{
\frac{\pi(\bx_i, 1; \etab)}{\phi(\bz_i; \bbeta, \etab)} -
\frac{I(d_i=1)}{f(1, \bz_i; \bbeta)}
\right\}\left\{
1 - f(0, \bz_i; \bbeta)
\right\}.
\eas
As the weights $1 - f(0, \bz_i; \bbeta)$ are always positive,
we abandon them and
propose testing $H_0$ based on
$U_s(\widehat\bbeta, \widehat\etab)$,
where
\bas
U_s(\bbeta, \etab) = \sum_{i=1}^m\left\{
\frac{\pi(\bx_i, 1; \etab)}{\phi(\bz_i; \bbeta, \etab)} -
\frac{I(d_i=1)}{f(1, \bz_i; \bbeta)}
\right\}
\eas
and $(\widehat\bbeta, \widehat\etab)$
is the maximum empirical likelihood estimator of $(\bbeta, \etab)$ under $H_0$.

\begin{theorem}\label{the:score-test}
Suppose that $\mathbb{E}[\{f(1, \bZ;\bbeta_0)\}^{-1}]<\infty$
and $\bSigma$ is positive definite. Then,
(a) $E\{U_s(\bbeta_0, \etab_0)\}\leq 0$ with the equality holding if and only
if $\omega_0 = 1$.
(b) Under $H_0: \omega=1$, as $N_0\rightarrow\infty$, the distribution of $N_0^{-1/2} U_s(\widehat\bbeta, \widehat\etab)$ converges to the normal distribution with mean zero and
variance $ \sigma_s^2$,
where $\sigma_{s}^2$ is defined in (8) of the supplementary information.
\end{theorem}

We can construct a consistent estimator of $\sigma_s^2$,
say $\widehat\sigma_s^2$,
in the same way that we constructed a consistent estimator for $\sigma^2$.
We define a score-like test statistic as
$
S = U_s(\widehat\bbeta, \widehat\etab)/ \{ \widehat N ^{1/2} \widehat\sigma_s\}$,
whose distribution converges to the standard normal distribution under $H_0$.
At the significance level $a$,
we reject the null hypothesis of $\omega=1$
if $S\leq Z_{a}$, where $Z_a$ is the $a$th quantile of the standard normal distribution.

\section{Simulation}\label{s:sim}

\subsection{Discrete-time capture--recapture experiments without one-inflation}
\label{sec:sim-dis}

We study the finite-sample performance of
the proposed two-step semiparametric empirical likelihood method
by comparing it with \cite{liu2020maximum}'s
one-step empirical likelihood method in
 discrete-time capture--recapture experiments without one-inflation.
We fix $N_0 = 200$ or 400 and
generate data for the following two scenarios:

\begin{description}

\item[A.]
The covariate is $\bZ = (1, X_1, X_2, Y)^\T$,
where $X_1\sim{\rm Bi}(1, 0.5)$ and
$X_2\sim{\rm Bi}(1, 0.7)$ are fully observed,
and $Y\sim {\rm U}(0, 1)$ is subject to missingness.
Given $\bZ = \bz$,
the number of captures $D$ is generated from a  Binomial  regression model~\eqref{eq:mod-bi} where $K = 17$ and
$\bbeta_0 = (-1.5, -0.3, -1.2, 0.5)^\T$.
Given $(X_1, X_2, D) = (x_1, x_2, k)$, the probability of observing $Y$ is
$\pi(x_1, x_2, k) = \{1 + \exp(0.3 - 0.5x_1 - 0.5x_2 - 0.5k)\}^{-1}$.

\item[B.]
The values are the same as those in Scenario A except that $X_2\sim {\rm U}(0,2)$.

\end{description}

In Scenario A,
the one-step empirical likelihood method is directly applicable because
the fully observed covariates $X_1$ and $X_2$
are both discrete.
However, it is not directly applicable to Scenario B,
which involves a fully observed continuous covariate.
To apply the one-step empirical likelihood method in Scenario B,
we transform $X_2$ to a binary variable $X_2^*$,
which is defined to be 0 if $X_2$ is less than its sample mean
and 1 otherwise.
The one-step empirical likelihood method is
then applied with $X_2^*$ in place of $X_2$.

We first compare the one- and two-step maximum empirical likelihood
estimators of $N$ and $\bbeta$.
Table~\ref{tab:sim-pint} gives their biases and
root-mean-square errors (RMSEs) based on 2000 simulation repetitions for each scenario.
We see that
the biases of the one- and two-step maximum empirical likelihood estimators
are both negligible in Scenario A.
The two-step maximum empirical likelihood estimators are comparable
or slightly inferior to the one-step maximum empirical likelihood estimators
in terms of mean square errors.
This is probably because the one-step empirical likelihood method
 makes full use of all data simultaneously,
whereas the two-step empirical likelihood method does this step by step, and
hence, has a potential loss of efficiency.

When the fully observed $X_2$ is a continuous variable, as in Scenario B,
the two-step maximum empirical likelihood estimators are not only nearly unbiased
but are also much more efficient
than the one-step maximum empirical likelihood estimators, which often have large biases.
This shows that the two-step maximum empirical likelihood estimators are more reliable and that,
with $X_2^*$ in place of $X_2$,
the resulting one-step empirical likelihood method can produce distorted point estimates.

\begin{table}[htbp]
\tabcolsep4.8pt
\centering
\caption{Simulated biases and
root-mean-square errors
of estimators of the abundance $N$ and the regression parameter
$\bbeta = (\beta_1,\beta_2,\beta_3,\beta_4)^\T$.
$N$ is rounded to the nearest integer.
$\beta_j$'s are multiplied by a thousand.}
\begin{tabular}{cccrrrrrrrrrr}
\toprule
 & & & \multicolumn{5}{c}{Scenario A} &
\multicolumn{5}{c}{Scenario B}
\\
 & $N_0$ & Method & $N$ & $\beta_1$ & $\beta_2 $ & $\beta_3$ & $\beta_4$
			 & $N$ & $\beta_1$ & $\beta_2 $ & $\beta_3$ & $\beta_4$
\\
\midrule
Bias &
200
 & One step & 1 & $-1$ & $-2$ & 6 & $-8$ & $-11$ & $-445$ & 7 & 126 & $-11$ \\
 & &Two step & 1 & $-6$ & $-4$ & $-1$ & $-2$ & 4 & -7 & -7 & $-10$ & 6
\\
 & 400 & One step & 1 & $-1$ & 0 & 0 & $-4$ & $-25$ & $-446$ & 11 & 128 & $-11$ \\
 & &Two step	 & 2 & $-3$ & $-1$ & $-3$ & 0 & 5 & $-8$ & $-1$ & $-2$ & 3\\
RMSE & 200
 & One step & 9 & 169 & 141 & 144 & 255
 & 17 & 495 & 173 & 243 & 331
\\
 & &Two step & 10 & 177 & 153 & 156 & 259
 & 17 & 232 & 177 & 205 & 313
\\
 & 400 & One step 	 & 13 & 122 & 101 & 104 & 182 &
29 & 470 & 127 & 196 & 232 \\
 & & Two step	 & 14 & 126 & 108 & 112 & 183 &
24 & 157 & 128 & 143 & 221\\
\bottomrule
\end{tabular}
\label{tab:sim-pint}
\end{table}

We next compare the scaled empirical likelihood ratio confidence interval
${\mathcal I}_{\rm SEL}$ and the one-step
empirical likelihood ratio confidence interval
${\mathcal I}_{\rm EL}$ of \cite{liu2020maximum}.
Table~\ref{tab:sim-CP} tabulates
the simulated coverage probabilities
of the one- and two-sided confidence intervals
of each method at the 95\% and 99\% levels.
The proposed confidence interval ${\mathcal I}_{\rm SEL}$
and the accompanying one-sided intervals have very close-to-nominal coverage probabilities in all scenarios.
The  one-step empirical likelihood ratio confidence interval ${\mathcal I}_{\rm EL}$
has desirable two-sided coverage accuracy in Scenario A,
but has severe undercoverage in Scenario B at both the nominal level 95\% and at 99\%.
In addition,
its lower limit often has severe overcoverage and
its upper limit often has severe undercoverage in Scenario B.
This may be caused by the large bias of the one-step maximum empirical likelihood estimator in this scenario.

\begin{table}[htbp]
\tabcolsep3.7pt
\centering
\caption{Simulated coverage probabilities (\%)
of each two-sided confidence interval and the lower and upper limits
at the 95\% and 99\% levels.
All numbers are rounded to the nearest integer.}
\begin{tabular}{ccccccccccccccc}
\toprule
 & &\multicolumn{6}{c}{Scenario A} &
\multicolumn{6}{c}{Scenario B} \\
 & &
\multicolumn{2}{c}{Two-sided} &
\multicolumn{2}{c}{Lower limit} &
\multicolumn{2}{c}{Upper limit} &
\multicolumn{2}{c}{Two-sided} &
\multicolumn{2}{c}{Lower limit} &
\multicolumn{2}{c}{Upper limit} \\
$N_0$ & Level &
95 & 99 & 95 & 99 & 95 & 99 & 95 & 99 & 95 & 99 & 95 & 99
\\
\midrule
200
 & ${\mathcal I}_{\rm EL}$ & 96 & 99 & 96 & 99 & 96 & 99 &
87 & 96 & 99 & 100 & 80 & 93
\\ & ${\mathcal
I}_{\rm SEL}$ & 96 & 99 & 96 & 99 & 95 & 99 &
95 & 99 & 95 & 99 & 95 & 99
\\
400
 & ${\mathcal I}_{\rm EL}$ & 94 & 99 & 94 & 98 & 95 & 99 &
78 & 92 & 100 & 100 & 68 & 87\\
 & ${\mathcal I}_{\rm SEL}$ & 94 & 99 & 94 & 99 & 95 & 99 &
94 & 99 & 94 & 99 & 94 & 98\\
\bottomrule
\end{tabular}
\label{tab:sim-CP}
\end{table}

\begin{figure}[htbp]
 \centering
 \includegraphics[width=4.8cm]{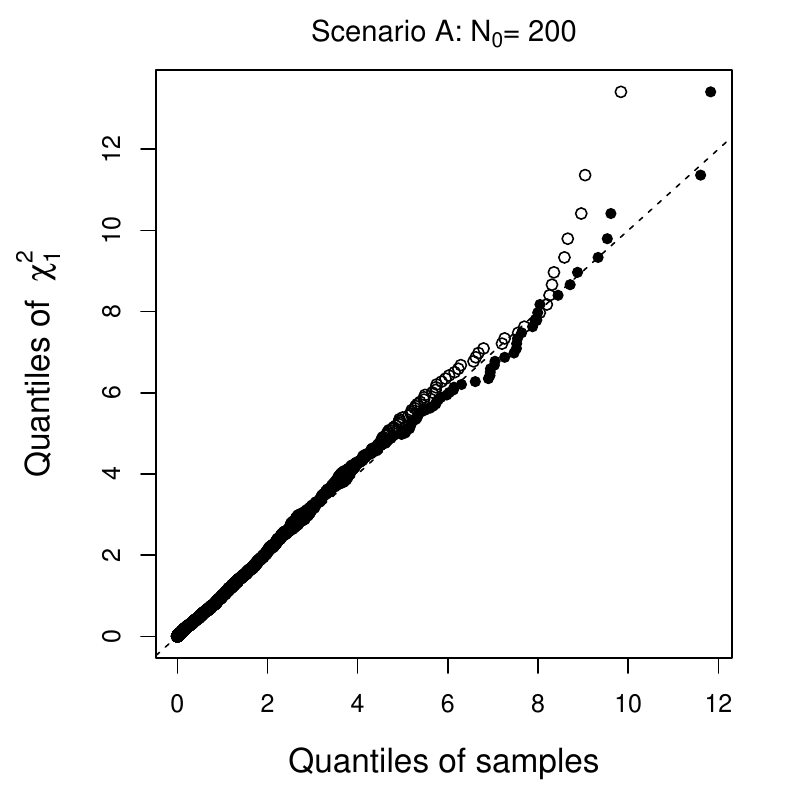}
 \includegraphics[width=4.8cm]{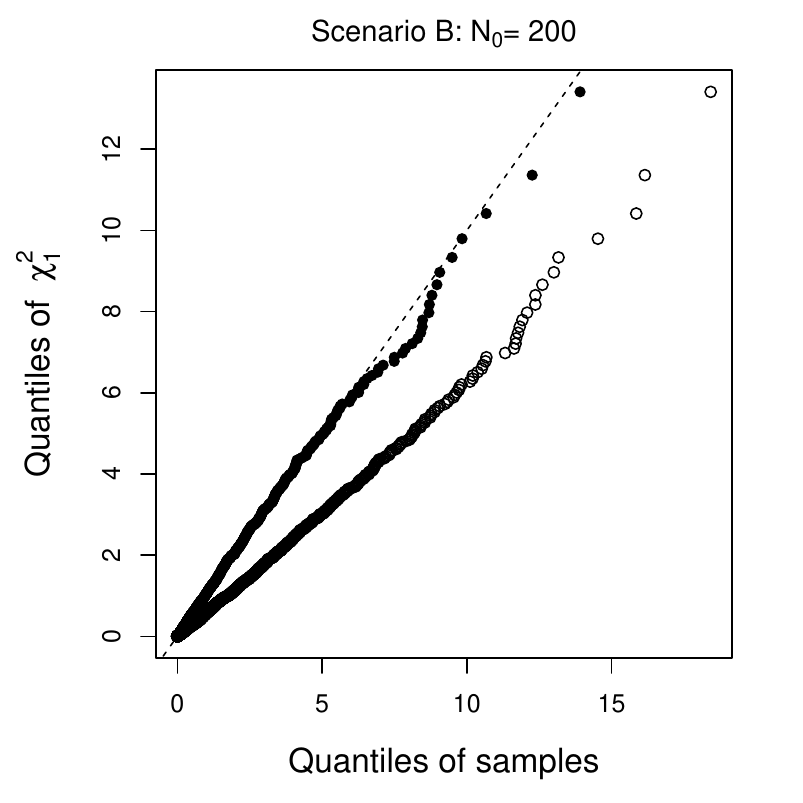} \\
 \includegraphics[width=4.8cm]{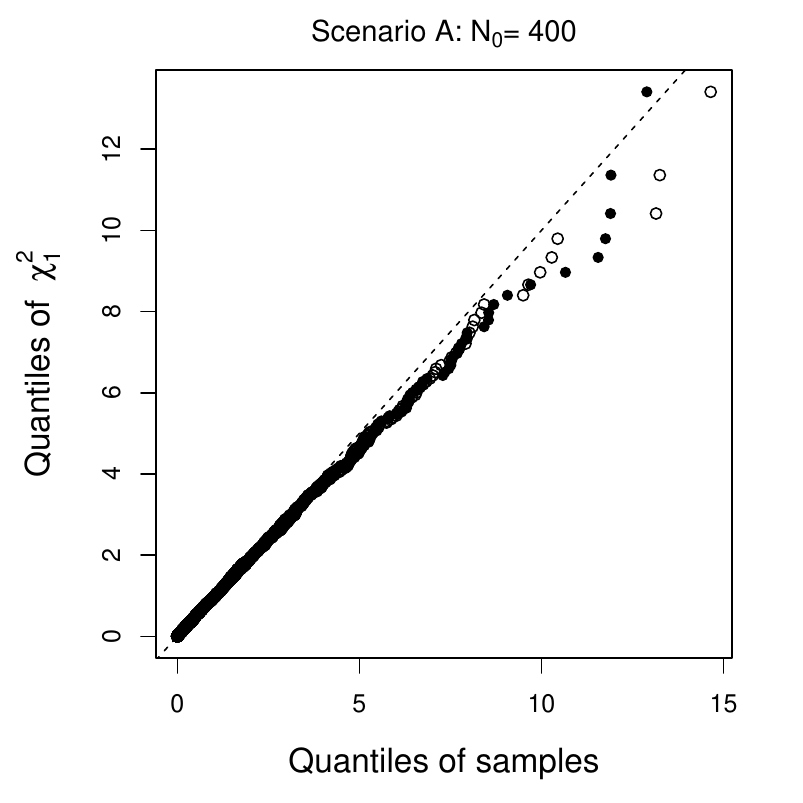}
 \includegraphics[width=4.8cm]{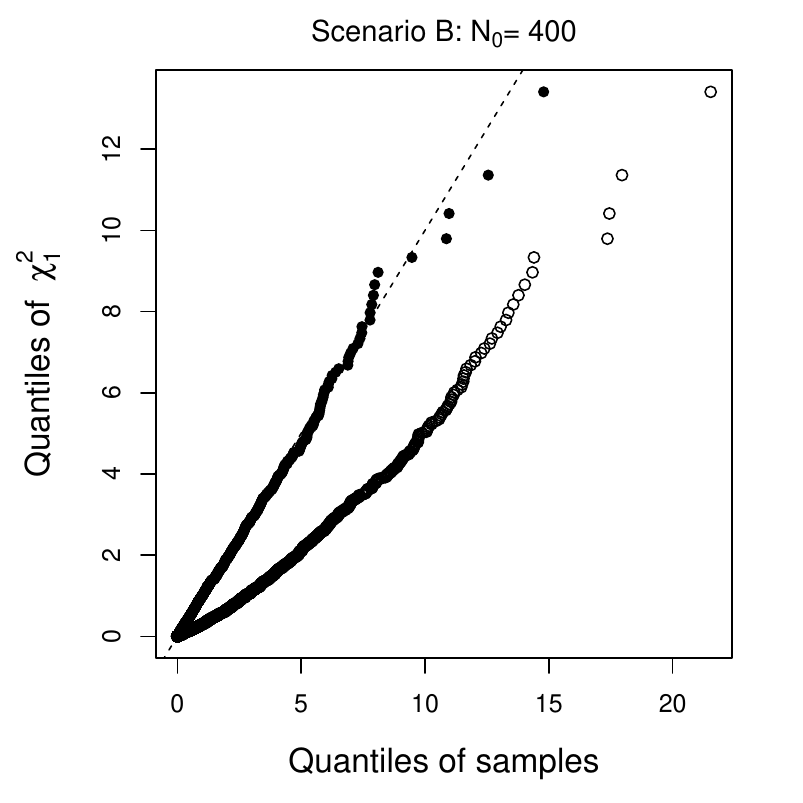}
 \caption{QQ plots of the empirical likelihood ratio test statistic ({\scriptsize$\circ$}) and
the scaled empirical likelihood ratio test statistic ($\bullet$) with
their limiting distribution in Scenarios A and B.
In both panels, the dashed line is the identity line.}\label{fig:elr}
\end{figure}

Finally, we check the approximation performance of
the limiting chi-square distribution
to the finite-sample distributions of
\cite{liu2020maximum}'s empirical likelihood ratio statistic and
our scaled empirical likelihood ratio statistic for testing whether $N=N_0$.
Figure~\ref{fig:elr} displays their
quantile--quantile (QQ) plots.
Clearly, the sampling distribution of the test statistic of our scaled empirical likelihood ratio is close to the limiting $\chi_1^2$ distribution
in both scenarios when $N_0=200$,
and the approximation accuracy is much better when $N_0$
increases to 400.
This explains why our two-step empirical likelihood ratio confidence intervals
always have desirable coverage probabilities.
The sampling distribution of \cite{liu2020maximum}'s
empirical likelihood ratio test statistic is also close to $\chi_1^2$ in Scenario A,
whenever $N_0=200$ or 400.
However, this is not the case in Scenario B.
This may explain why the one-step empirical likelihood ratio confidence intervals
have desirable coverage probabilities in Scenario A but have large coverage errors in Scenario B.

\subsection{Continuous-time capture--recapture experiments without one-inflation}
\label{sec:sim-con}

\cite{liu2020maximum}'s one-step empirical likelihood method works
 only if the total number of captures is finite.
In continuous-time capture--recapture studies,
the number of captures may be infinite, so the one-step empirical likelihood method fails.
The proposed two-step empirical likelihood method does not have such a limitation,
although it involves an infinite summation.
To facilitate the practical use of our two-step empirical likelihood method,
we suggest approximating $\phi_e(\bz; \bbeta)$ by
$$
\phi_e(\bz; \bbeta)\approx
\sum_{k=k_{\rm min}}^{k_{\rm max}}
\pi(\bx, k) \frac{\lambda_*^k
e^{-\lambda_*}}{k!},\quad
$$
where $\lambda_* = \lambda(\bz; \bbeta)$, and
$k_{\rm min}$ and $k_{\rm max}$ are related to $\lambda_*$ via
\[
k_{\rm min} =
\begin{cases}
1, & \lambda_*\leq16, \\
\max(1, \lfloor\lambda_* -
5 \lambda_*^{1/2} \rfloor), & \lambda_*>16,
\end{cases} \quad
k_{\rm max} =
\begin{cases}
30, & \lambda_*\leq16, \\
\lceil\lambda_* +
5\lambda_*^{1/2} \rceil, & \lambda_*>16.
\end{cases}
\]
Here, $\lfloor x\rfloor$ and $\lceil x\rceil$, respectively,
represent the floor and ceiling functions of $x$.
Our simulations show that
this approximation works very well.
To study the finite-sample performance of
the two-step empirical likelihood method
for continuous-time capture--recapture data,
we generate data for the following scenario.

\begin{description}
\item[C.]
The values are the same as those in Scenario A except that
the number of captures $D$ for a given $ \bz$ is generated from a
Poisson regression model~\eqref{eq:mod-po} with
$\bbeta_0 = (0.5, -0.3, -1.2, 0.5)^\T$.
\end{description}

Table~\ref{tab:sim-continuous} presents the simulated biases
and RMSEs of
the two-step maximum empirical likelihood estimators of
$N$ and $\bbeta = (\beta_1, \beta_2, \beta_3, \beta_4)^\T$,
as well as the coverage probabilities of the scaled empirical likelihood ratio
confidence interval ${\mathcal I}_{\rm SEL}$
at the 95\% and 99\% levels when data were generated for Scenario C.
We see that the two-step maximum empirical likelihood estimator of
the regression parameter has negligible bias,
and its RMSE decreases as $N_0$ increases, as expected.
For abundance estimation,
the relative bias (bias divided by $N_0$) and
the relative RMSE (RMSE divided by $N_0$) of
the proposed abundance estimator
both decrease as $N_0$ increases from 200 to 400.
The proposed confidence interval ${\mathcal I}_{\rm SEL}$
always has very accurate one- and two-sided coverage probabilities.

\begin{table}[htbp]
\tabcolsep2.4pt
\centering
\caption{Simulated biases and root-mean-square errors (unit: 0.001)
of the two-step maximum empirical likelihood estimators of $N$
and $\bbeta = (\beta_1,\beta_2,\beta_3,\beta_4)^\T$,
as well as the simulated coverage probabilities (\%)
of the two-sided confidence interval and lower and upper limits of
${\mathcal I}_{\rm SEL}$ at the 95\% and 99\% levels
when data were generated for Scenario C.}
\begin{tabular}{ccccccccccccccccc}
\toprule
 & \multicolumn{5}{c}{Bias} &
\multicolumn{5}{c}
{RMSE} &
\multicolumn{2}{c}
{Two-sided} &
\multicolumn{2}{c}
{Lower limit} &
\multicolumn{2}{c}
{Upper limit} \\
\midrule
$N_0$ & $ N$ & $\beta_1$ & $\beta_2 $ & $\beta_3$ & $\beta_4$ &
$ N$ & $\beta_1$ & $\beta_2 $ & $\beta_3$ & $\beta_4$ &
95 & 99 & 95 & 99 & 95 & 99
\\
200 & 
1E4\tnote{1} & $-32$ & $-11$ & $-11$ & 10 &
3E4 & 290 & 248 & 294 & 420 &
95 & 99 & 94 & 99 & 97 & 99\\
400 & 
1E4 & $-15$ & $-4$ & $-9$ & 5 &
4E4 & 200 & 171 & 205 & 295 &
94 & 99 & 94 & 99 & 95 & 99\\
\bottomrule
\end{tabular}
\label{tab:sim-continuous}
\begin{tablenotes}
\item[1] 1E4, $10^{4}$; 3E4, $3 \times 10^{4}$; 4E4, $4 \times 10^{4}$.
\end{tablenotes}
\end{table}

\subsection{Capture--recapture experiments with one-inflation}

We use the following scenario to represent capture--recapture experiments with one-inflation.

\begin{description}
\item[D.] As Scenario A except that
the number of captures $D$ is generated from a one-inflated  Binomial  regression model
\eqref{eq:cap-prob-inflate}.
\end{description}

We first examine the type I error and power of the score-like test $S $
for testing $H_0 : \omega = 1$ under model~\eqref{eq:cap-prob-inflate}.
We consider eight values of $\omega_0$ ranging from 0.3 to 1 with steps of 0.1.
Table~\ref{tab:sim-oi-test} presents
the rejection rates
based on 50,000 repetitions for $\omega_0 = 1$
and 2000 repetitions for $\omega_0 < 1$.
When $\omega_0 = 1$, the rejection rates of $S$ are its type I errors.
They are close to or slightly less
than the nominal significance levels at the 5\% and 10\% nominal levels,
and are slightly inflated at the 1\% significance level.
This implies that the proposed score-like test has good control of its type I error,
and that the limiting chi-square distribution provides
a good approximation to the sampling distribution of $S$ when $N_0$ is large.
When $\omega_0 < 1$, the null hypothesis is violated
and the rejection rates are all powers.
We see that the score-like test $S$ has desirable and increasing power as
$\omega_0$ decreases from 1 to 0.5.
The power slightly decreases as $\omega_0$ decreases further from 0.5 to 0.3.
A possible reason is that score tests are local tests, and they have good performance
for alternatives that are close to the null hypothesis.
However, their performance is not guaranteed if the alternative is far from the null hypothesis.

\begin{table}[htbp]
\tabcolsep3pt
\centering
\caption{Rejection rates (\%) of the proposed score-like test for $H_0: \omega=1$
at the significance levels 10\%, 5\%, and 1\%.}
\begin{tabular}{cccccccccc}
\toprule
Level & $N_0$ & $\omega_0=1$ & $\omega_0=0.9$ & $\omega_0=0.8$ &
$\omega_0=0.7$ & $\omega_0=0.6$ & $\omega_0=0.5$ &
$\omega_0=0.4$ & $\omega_0=0.3$ \\
\midrule
10\% & 200
 & 8.72 & 54.00 & 88.40 & 96.00 & 99.65 & 99.95 & 99.90 & 98.85 
\\
 & 400
 & 9.32 & 77.25 & 98.75 & 99.95 & 100.0 & 100.0 & 100.0 & 100.0 
\\
5\% & 200
 & 4.99 & 42.85 & 81.50 & 93.70 & 99.30 & 99.70 & 99.85 & 98.70 
\\
 & 400
 & 5.15 & 67.55 & 97.50 & 99.95 & 100.0 & 100.0 & 100.0 & 100.0 
\\
1\% & 200
 & 1.57 & 23.95 & 62.30 & 84.85 & 96.70 & 99.00 & 98.40 & 96.95 
\\
 & 400
 & 1.49 & 45.65 & 92.50 & 99.45 & 100.0 & 100.0 & 100.0 & 100.0 
\\
\bottomrule
\end{tabular}
\label{tab:sim-oi-test}
\end{table}

Next, we compare the performance of the
two-step maximum empirical likelihood estimators $\widehat N_e$
and $\widehat N$ with and without the one-inflation assumption.
We consider three values, 0.9, 0.7, and 0.5, for $\omega_0$ in Scenario D,
and, in Table~\ref{tab:sim-oi-est}, we report the biases and RMSEs of these two estimators based on
2000 repetitions.
Our general finding is that ignoring one-inflation results in positive bias
and a larger RMSE.
The largest bias and RMSE of $\widehat N$ occurred when $\omega_0=0.5$ and $N_0 = 400$. These were about
 18 and 6 times larger than the values for $\widehat N_e$.

\begin{table}[htbp]
\tabcolsep7.5pt
\centering
\caption{Simulated biases and
root-mean-square errors
of the two-step maximum empirical likelihood estimators $\widehat N$
and $\widehat N_e$ in Scenario D.
All numbers are rounded to the nearest integer.}
\begin{tabular}{ccccccccccccc}
\toprule
 & \multicolumn{6}{c}{$N_0 = 200$} & \multicolumn{6}{c}{$N_0 = 400$}
\\
 & \multicolumn{2}{c}{$\omega_0 = 0.9$}
 & \multicolumn{2}{c}{$\omega_0 = 0.7$}
 & \multicolumn{2}{c}{$\omega_0 = 0.5$}
 & \multicolumn{2}{c}{$\omega_0 = 0.9$}
 & \multicolumn{2}{c}{$\omega_0 = 0.7$}
 & \multicolumn{2}{c}{$\omega_0 = 0.5$}
\\
\midrule
 & $\widehat N$ & $\widehat N_e$
 & $\widehat N$ & $\widehat N_e$
 & $\widehat N$ & $\widehat N_e$
 & $\widehat N$ & $\widehat N_e$
 & $\widehat N$ & $\widehat N_e$
 & $\widehat N$ & $\widehat N_e$
\\
Bias & 9 & 3 & 32 & 5 & 80 & 8 &
16 & 3 & 59 & 5 & 145 & 8
\\
RMSE & 13 & 11 & 32 & 13 & 80 & 17 &
20 & 15 & 59 & 18 & 145 & 23
\\
\bottomrule
\end{tabular}
\label{tab:sim-oi-est}
\end{table}

Lastly, we compare the proposed empirical likelihood ratio
confidence intervals
${\mathcal I}_{{\rm SEL}e}$ and
${\mathcal I}_{\rm SEL}$
with and without the one-inflation assumption.
 Table~\ref{tab:sim-oi-interval}
 presents their coverage probabilities at the 95\% and 99\% levels.
When ignoring one-inflation,
the two-sided confidence intervals and the lower limits of
${\mathcal I}_{\rm SEL}$ often produce severe undercoverage, whereas the upper limits often produce severe overcoverage, especially when $\omega_0$ is as low as 0.5.
This is probably caused by the large positive bias of
$\widehat N$ when one-inflation is mistakenly ignored.
In contrast, the confidence interval
${\mathcal I}_{{\rm SEL}e}$ always
has very accurate one- and two-sided coverage probabilities.

\begin{table}[htbp]
\tabcolsep7.8pt
\centering
\caption{Simulated coverage probabilities (\%)
of each two-sided confidence interval and the lower and upper limits of
${\mathcal I}_{{\rm SEL}}$ and
${\mathcal I}_{{\rm SEL}e}$ in Scenario D.}
\begin{tabular}{cccccccccccc}
\toprule
 & & & \multicolumn{2}{c}{$\omega_0 = 0.9$}
 & \multicolumn{2}{c}{$\omega_0 = 0.7$}
 & \multicolumn{2}{c}{$\omega_0 = 0.5$} \\
 & Level & $N_0$ & ${\mathcal I}_{{\rm SEL}}$ & ${\mathcal I}_{{\rm SEL}e}$
 & ${\mathcal I}_{{\rm SEL}}$ & ${\mathcal I}_{{\rm SEL}e}$
 & ${\mathcal I}_{{\rm SEL}}$ & ${\mathcal I}_{{\rm SEL}e}$
\\
\midrule
Two-sided
 & 95 & 200
 & 92 & 95 & 62 & 95 & 13 & 94
\\
 & &400
 & 91 & 95 & 37 & 95 & 1 & 95
\\
 & 99 & 200
 & 98 & 99 & 85 & 98 & 32 & 98
\\
 & &400
 & 98 & 99 & 61 & 99 & 4 & 99
\\
Lower limit
 & 95 & 200
 & 88 & 94 & 49 & 95 & 8 & 94
\\
 & &400
 & 82 & 96 & 26 & 94 & 0 & 94
\\
 &99  & 200
 & 97 & 99 & 77 & 99 & 24 & 99
\\
 & &400
 & 96 & 99 & 52 & 99 & 2 & 99
\\
Upper limit
 & 95 & 200
 & 98 & 95 & 100 & 95 & 100 & 96
\\
 & &400
 & 99 & 95 & 100 & 96 & 100 & 96
\\
 & 99 & 200
 & 99 & 98 & 100 & 98 & 100 & 98
\\
 & &400
 & 100 & 99 & 100 & 99 & 100 & 99
\\
\bottomrule
\end{tabular}
\label{tab:sim-oi-interval}
\end{table}

\section{Real example}
\label{s:data}
We apply the proposed two-step semiparametric empirical likelihood method
to the yellow-bellied prinia data collected in Hong Kong.
Altogether, 163 distinct birds were captured at least once
during the 17 weeks from January to April 1993.
Following \cite{liu2020maximum},
we consider three covariates: fat index $X_1$,
wing length $X_2$, and tail length $Y$.
The covariates $X_1$ and $X_2$ are fully observed variables,
whereas $Y$ is subject to missingness with missing rate 25\%.
The covariate $X_1$ is binary,
being 1 for fat and 0 for not fat.
$X_2$ is a continuous variable
ranging from 43 to 49\,mm.

To apply the one-step empirical likelihood method,
\cite{liu2020maximum} suggested
transforming $X_2$ to a binary variable $X_2^*$,
equal to $1$ if the wing length is above 45.5\,mm and 0 otherwise.
Instead of using the surrogate $X_2^*$, we take
the original continuous variable $X_2$ as a covariate.
Using the estimation method in Section~\ref{sec:step-one},
we fitted the non-missingness probability model based on the prinia data:
\bas
\pi(x_1, x_2, k; \widehat\eta) = \frac{
\exp(-3.71+ 0.84x_1 + 0.06x_2 + 1.44k)}{
1 + \exp(-3.71+ 0.84x_1 + 0.06x_2 + 1.44k)}.
\eas
When testing for one-inflation under model~\eqref{eq:cap-prob-inflate},
the proposed score-like test statistic is
$S = -1.04$ with $p = 14.9$\%.
At the 5\% significance level,
this indicates that there is not enough evidence
to support the existence of one-inflation for the number of captures.
Under the Huggins--Alho model~\eqref{eq:mod-bi} without one-inflation,
the two-step maximum empirical likelihood estimate for the abundance is
733 (standard error 240)
and  the  two-step empirical likelihood ratio  confidence interval is [436, 1717]
at the 95\% level.
In contrast, the one-step maximum empirical likelihood estimate of $N$ is 740 (standard error 217)
and the one-step empirical likelihood ratio confidence interval is [452, 1652].
Because the transformation of $X_2$ to $X_2^*$ may lose information,
we believe that the results of the two-step empirical likelihood method
are more reliable than those of the one-step empirical likelihood method.

\section{Conclusion and discussion}
\label{s:con}

In the context of discrete-time capture-recapture studies
when covariates are missing at random,
the maximum likelihood estimation method
proposed by \cite{liu2020maximum} is recommended when the
fully-observed covariates are all categorical variables.
However, it is not directly applicable to the
continuous-time capture-recapture data and the case
where some fully-observed covariates take continuous values.
In this article, we allow for the fully-observed
continuous covariates and
propose a two-step semiparametric EL abundance  method for abundance estimation
under Binomial and Poisson regression models and associated one-inflated models.
Theoretically, we show that
the two-step maximum EL estimators follow
normal limiting distributions and
the scaled EL ratio statistics of the abundance
have a  chisquare limiting distribution.
A score-like test for the one inflation parameter is also derived.

In the development of our two-step empirical likelihood method,
we imposed a parametric model on
the non-missingness probability function $\pi(\bx, k)$.
To alleviate the possible risks of model misspecification,
one may make a nonparametric model assumption and
estimate $\pi(\bx, k)$ by nonparametric techniques.
Without loss of generality, we write the fully observed covariate
$\bx$ as $(\bu^{\T}, \rv_{1}, \ldots, \rv_{q})^\T$,
where the elements of $\bu$ are
categorical and the $\rv_i$'s are continuous.
Similarly, we write the observation $\bx_i$ as
$(\bu_i^{\T}, \rv_{i1}, \ldots, \rv_{iq})^\T$ for $i = 1,\ldots, n$.
A nonparametric estimator of $\pi(\bx, k)$ is
\bas
\frac{\sum_{i=1}^m
I(\bu_i = \bu, d_i = k)\prod_{j=1}^q K_h(\rv_{ij} - \rv_j)}
{\sum_{i=1}^n
I(\bu_i = \bu, d_i = k)\prod_{j=1}^q K_h(\rv_{ij} - \rv_j)},
\eas
where $K_h(\cdot) = K(\cdot/h)$,
$K(\cdot)$ is a kernel function, and $h$ a bandwidth.
Our two-step semiparametric empirical likelihood method still works
with the above estimator in place of $\pi(\bx, k)$.
However, this method may suffer from the curse of dimensionality
when $q$ is large.

\backmatter

\bmhead{Supplementary information}
Supplementary information available 
online
contains the proofs of Theorems \ref{prop:eta}--\ref{the:score-test} and
Corollary \ref{col:discrete}.

\bmhead{Acknowledgments}
The authors would like to thank Dr~W.~H.~Hwang for providing the prinia data.
This research is supported by
the National Key R\&D Program of China (2021YFA1000100 and 2021YFA1000101),
the National Natural Science Foundation of China (12101239, 12171157, 11831008, 11971171, 12171310,
32030063, and 71931004),
the Natural Sciences and Engineering Research Council of Canada (RGPIN-2020-04964),
the China Postdoctoral Science Foundation (2020M681220),
the 111 project (B14019),
and the Basic Research Project of Shanghai Science and Technology Commission (22JC1400800).

\end{document}